\documentclass[ aps,%
floatfix,%
final,%
notitlepage,%
oneside,%
onecolumn,%
nobibnotes,%
nofootinbib,%
superscriptaddress,%
showpacs,%
]%
{revtex4}

\usepackage{epsfig}
\usepackage{amsfonts}
\usepackage{amsmath}
\usepackage{epsfig}
\usepackage{graphics}

\begin{document}

\title{Light hadron production in inclusive $pp$-scattering at LHC}

\author{Likhoded A.K.}\email{Anatolii.Likhoded@ihep.ru}
\affiliation{Institute for High Energy Physics, Protvino, Russia}
\author{Luchinsky A.V.}\email{Alexey.Luchinsky@ihep.ru}
\affiliation{Institute for High Energy Physics, Protvino, Russia}
\author{Novoselev A.A.}\email{Alexey.Novoselov@cern.ch}
\affiliation{Institute for High Energy Physics, Protvino, Russia}
\affiliation{Moscow Institute of Physics and Technology, Dolgoprudny, Russia}

\begin{abstract}
The inclusive production of light mesons in $pp$-scattering is considered
in the framework of reggeon phenomenology with supercritical Pomeron.
Available low-energy data can be explained with three reggeon particles
taken into account. With the results obtained rapidity and pseudorapidity
distributions for light-meson production at the LHC energies are predicted.
\end{abstract}


\maketitle

\section{Introduction}

It is well known that the growth of total cross sections can be explained
in the framework of reggeon theory with supercritical Pomeron \cite{Baker76}.
In this approach the amplitude of an elastic hadron-hadron scattering
is written down as a sum of reggeon diagrams which corresponds to multiple rescattering.
The first term of such eikonal expansion is the contribution
of a ``bare'' Pomeron:

\begin{eqnarray}
A\left(s,t\right) & = & \left(i-\cot\frac{\pi\alpha(t)}{2}\right)s^{\alpha(t)}\gamma(t),\end{eqnarray}
where the Regge trajectory in the $j$-plane is parameterized
by a linear function $\alpha(t)=1+\Delta+\alpha't$ with the intercept
$\alpha(0)=1+\Delta$.

According to the perturbative QCD calculations \cite{Lipatov} the vacuum
exchange singularity has rather complex form. It can be expressed
as a sum of poles in the angular momentum plane with $1\le j\le1+\Delta$,
where

\begin{eqnarray*}
\Delta & < & \frac{12 \alpha_{s} \ln 2 }{\pi}.\end{eqnarray*}
For $\alpha_{s}=0.2\div0.25$ this
allows $\Delta$ to approach rather large
value of $0.55\div0.65$.

The information on the parameter $\Delta$ can  also be obtained from the
analysis of deep inelastic scattering structure functions in the region
of small Bjorken variable $x$ \cite{Gribov}, where the Regge asymptotic
is valid. The power parameter $\Delta$, defined as $\Delta=\partial\ln f/\partial(1/x)-1$,
varies from $\Delta=0.23$ at $Q^{2}=16\,\mathrm{GeV}^{2}$ to $\Delta=0.5$
at $Q^{2}=400\,\mathrm{GeV}^{2}$.

The data on total cross sections were analyzed using eikonal model with different
values of parameter $\Delta$. In the earlier works \cite{Capella}
total cross sections were fitted using $\Delta\sim0.13$. In the paper
\cite{Kopelevich} a double-pole approximation for bare Pomeron
was used for the first time with $\Delta$ being in the region $0.09<\Delta<0.3$.
In the recent work \cite{Petrov02} the triple-pole approximation
was used with $\Delta_{1}  =  0.058, \Delta_{2}  =  0.0167, \Delta_{3}  =  0.203$.

Other approach to determine the parameters of bare
Pomeron is connected with the inclusive production in the central region \cite{Kobr,Chliap,Likhoded91}.
The argument in support of such method is the AGK cancellation of
cut contributions in this region \cite{Abram}. 
Therefore
one deals only with the first term --- double bare Pomeron diagram.
As a result, the asymptotic of inclusive production cross section
in the central region has a simple form:

\begin{eqnarray}
\left.\frac{d\sigma}{dy}\right|_{y=0} & \sim & s^{\Delta}\label{eq:opt}.
\end{eqnarray}

As it was mentioned earlier, only the leading term is taken into account
in the expression (\ref{eq:opt}), while the contributions of secondary trajectories
are neglected. However, at low and moderate energies ($\sqrt{s}<100$ GeV)
the growth of particle density is caused by
contributions of secondary trajectories in Mueller-Kanchelli diagrams \cite{Mueller}. These
terms are also important at higher energies for large values of
rapidity $y\sim\ln x+\ln\left(\sqrt{s}/m_{\perp}\right)$. The analysis
of experimental data on $d\sigma/dy|_{y=0}$ presented in our previous
works \cite{Chliap,Likhoded91} shows that power behavior of this
inclusive cross section is caused by the Pomeron singularity with $\Delta\approx0.17$.
In that analysis only low-energy data with $\sqrt{s} \leq 900$ GeV were
used. In the present work we are going to check whether new measurements
of cross-sections of inclusive production of charged particles
and $K_{S}$-mesons, obtained at LHC at
$\sqrt{s}=0.900$, $2.36$ and $7$ TeV energies agree with predictions
based on the low-energy data analysis.

\section{Double-Reggeon Exchange}

According to the generalized optical theorem and factorization of the
leading regge singularities the inclusive cross section
of the reaction $ab\to c+X$
in the central
region can be written down as
\begin{eqnarray}
E_{c}\frac{d^{3}\sigma}{d^{3}p_{c}} & = & \frac{1}{\pi}\frac{d^{2}\sigma}{dm_{T}^{2}dy} =
\frac{1}{s}\sum_{i,j}f_{ij}\left(m_{T}\right)\left(\frac{-t}{s_{0}}\right)^{\alpha_{i}(0)}\left(\frac{-u}{s_{0}}\right)^{\alpha_{j}(0)},
\label{eq:dist}\end{eqnarray}
where $i,j=R,P$ and $F$ for Reggeon, Pomeron and Froissaron exchange respectively. The intercept parameters for these particles are
\begin{eqnarray*}
  \Delta_R &=& -0.5,\quad \Delta_P=0.06, \quad \Delta_F=0.17.
\end{eqnarray*}
The kinematical variables are defined as $t=\left(p_{a}-p_{c}\right)^{2}\approx-\sqrt{s}m_{T}e^{-y}$,
$u=\left(p_{b}-p_{c}\right)^{2}\approx-\sqrt{s}m_{T}e^{y}$, $s=\left(p_{a}+p_{b}\right)^{2}$,
$m_{T}=\sqrt{m_{c}^{2}+p_{T}^{2}}$, $s_{0}=1\,\mathrm{GeV}^{2}$,
and rapidity $y$ is defined as
\begin{eqnarray*}
y & = & \ln\frac{E_{c}+p_{L}}{m_{T}},
\end{eqnarray*}
where $E_{c}$, $p_{L}$ and $p_{T}$ are energy, longitudinal and
transverse momentum of particle $c$ in the laboratory frame. The
dependence on transverse momentum is contained in the vertex functions
$f_{ij}(m_{T})$ which are determined from the experimental data. In our
work the following parametrization for these vertex functions is used:

\begin{eqnarray*}
f_{ij}\left(m_{T}\right) & = & A_{ij}\varphi_{ij}\left(m_{T}\right),\end{eqnarray*}
where constants $A_{ij}$ are determined from the fit of low-energy data ($\sqrt{s} \leq 900$ GeV)
on cross-sections of inclusive $pp\to c+X$ reactions in the central
region. Transverse momentum dependence of $\varphi_{ij}(m_{T})$-functions
can not be determined from Regge phenomenology, so, following the work
\cite{Kobr}, a simple exponential parametrization is used in our article:

\begin{eqnarray}
\varphi_{ij}\left(m_{T}\right) & = & \frac{\beta_{ij}^{2}}{2\left(\beta_{ij}m_c+1\right)}e^{-\beta_{ij}\left(m_{T}-m_c\right)}.\label{eq:phi}\end{eqnarray}
This functions obey a normalization condition\begin{eqnarray}
\int\limits _{m^{2}}^{\infty}\varphi_{ij}\left(m_{T}\right)dm_{T}^{2} & = & 1.\label{eq:norm}\end{eqnarray}

Parameter $\beta_{PP}$ for Pomeron singularity was determined from
the fit of low-energy data and equals

\begin{eqnarray*}
\beta_{PP} & = & 6\,\mathrm{GeV}^{-1}.
\end{eqnarray*}

For $\Delta=0.17$ singularity, which dominates at high energies,
$p_{T}$-spectrum of charge particles at LHC energies was used to obtain

\begin{eqnarray*}
\beta_{FF} & = & 4.7\,\mathrm{GeV}^{-1}.
\end{eqnarray*}

The decrease of this slope parameter with the growth of $\Delta$
can be connected with the decrease of the effective radius of the
Pomeron singularity, observed, for example, in the description of total
and elastic cross sections in the three-Pomeron model \cite{Petrov02}.

Let us first consider rapidity distributions at various energies squared
$s$. Integrating the expression (\ref{eq:dist}) over the transverse
mass one obtains

\begin{eqnarray*}
\frac{d\sigma}{dy} & = & \frac{1}{s_0}\sum_{i,j}A_{ij}\left({s \over s_0}\right)^{\left(\Delta_{i}+\Delta_{j}\right)/2}\cosh\left[\left(\Delta_{i}-\Delta_{j}\right)y\right].
\end{eqnarray*}
It is clear that this distribution does not depend on slope parameters
$\beta_{ij}$. It is important to note, that the in presented expression
$y$- and $\sqrt{s}$-dependencies are strongly
correlated. Thus the rapidity distributions at fixed $\sqrt{s}$ are as
important as the $\sqrt{s}$-dependence at fixed $y$.

Experimental data on these distributions and references
to original works are given in \cite{Chliap} (see references {[}10-27{]}
in that paper). Using this data we obtain the values of parameters
$A_{ij}$, presented in Tab. \ref{tab:Aij}.

\begin{table*}
\begin{centering}
\begin{tabular}{|c|c|c|c||c|c|c|c|}
\hline
\multicolumn{4}{|c||}{$pp\to c^{-}+X$} & \multicolumn{4}{c|}{$pp\to K_{S}+X$}\tabularnewline
\hline
$i/j$  & $R$  & $P$  & $F$ & $i/j$  & $R$  & $P$  & $F$\tabularnewline
\hline
\hline
$R$  & $-81.8$  & $67.9$ & $-53.1$ & $R$  & $11.6$ & $-11.5$ & $4.37$\tabularnewline
\hline
$P$  & $67.9$ & $6.39$ & $-0.741$ & $P$  & $-11.5$ & $12.5$ & $-6.09$\tabularnewline
\hline
$F$  & $-53.1$ & $-0.741$ & $9.16$ & $F$  & $4.37$ & $-6.09$ & $3.88$\tabularnewline
\hline
\end{tabular}
\par\end{centering}
\caption{Numerical values of parameters $A_{ij}$ for $c^{-}$ and $K_{S}$
production in inclusive $pp$-scattering\label{tab:Aij}}
\end{table*}

\begin{figure*}
\begin{centering}
\includegraphics{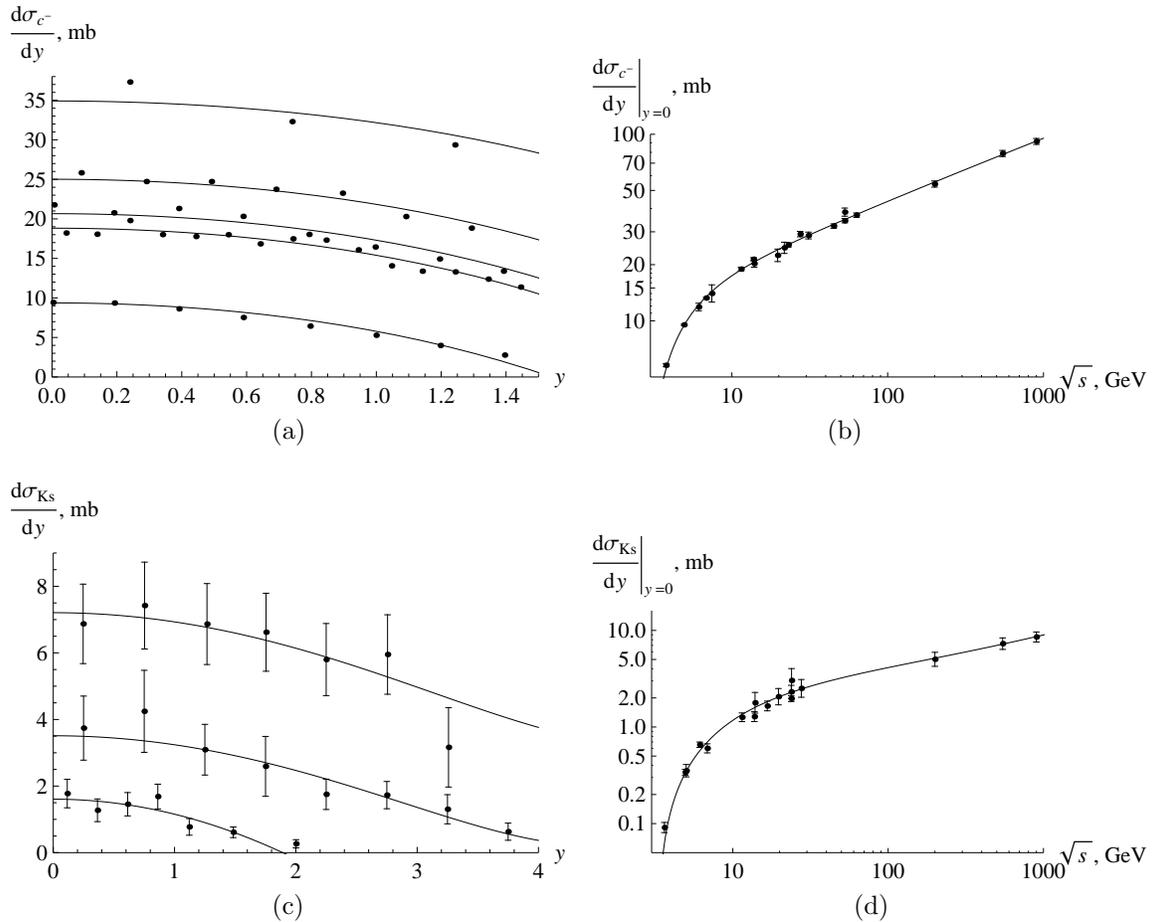}
\par\end{centering}
\caption{Fits of available experimental data by double-Reggeon exchange model:\\
(a) --- rapidity distributions of $pp\to c^{-}+X$ reaction cross
section. Curves and data sets in the figure correspond to energies
$\sqrt{s}=4.9$ GeV, 11.5 GeV, 13.9 GeV, 21.7 GeV and 53 GeV from
bottom to top; (b) --- energy dependence of differential cross section
of the reaction $pp\to c^{-}+X$ at $y=0$; (c) --- rapidity distributions
of $pp\to K_{S}+X$ reaction cross section at $\sqrt{s}=13.76$ GeV, 63 GeV and 546 GeV;
(d) --- energy dependence of the differential cross section of the
reaction $pp\to K_{S}+X$ at $y=0$}
\label{fig:dsdy}       
\end{figure*}

In Fig. \ref{fig:dsdy} rapidity distributions of inclusive
production of charged particles (upper figure) and $K_{S}$-meson (lower
figure) in proton-proton interactions at different energies and energy
dependence of $d\sigma/dy$ cross section at $y=0$ are given. Dots in these
figures represent experimental data used in the fitting procedure. Theoretical
description, obtained in the framework of Regge model with the parameters $A_{ij}$
presented above, are shown with lines.

Experimental data on charged particles distributions
at LHC energies are usually reported as distributions
over pseudorapidity

\begin{eqnarray*}
\eta & = & \frac{1}{2}\ln\frac{1+\cos\theta}{1-\cos\theta}=\ln\cot\frac{\theta}{2},
\end{eqnarray*}
where $\theta$ is scattering angle of the final particle. The
relation that connects rapidity and pseudorapidity is

\begin{eqnarray*}
y & = & \frac{1}{2}\ln\frac{\sqrt{m^{2}+p_{T}^{2}\cosh^{2}\eta}+p_{T}\sinh\eta}{\sqrt{m^{2}+p_{T}^{2}\cosh^{2}\eta}-p_{T}\sinh\eta}.
\end{eqnarray*}
It is clear, that in the limit of zero mass one has $\eta=y$.

To transfer from rapidity to pseudorapidity distribution
the information
on the $m_{T}$-dependence of vertex functions $\varphi_{ij}$ is needed:

\begin{eqnarray*}
\frac{d\sigma}{d\eta} & = & \int\limits _{m^{2}}^{\infty}\frac{p}{E}\frac{d^{2}\sigma}{dydm_{T}^{2}}dm_{T}^{2}.
\end{eqnarray*}
It was earlear mentioned, that Regge theory gives no information about
the form of these functions.
So the exponential parametrization (\ref{eq:phi}) which
 agrees with the experimental data is used.
In addition, we assume factorization of the slope parameters of these
functions, so that the following relation is fulfilled:

\begin{eqnarray*}
\beta_{ij} & = & \beta_{i}+\beta_{j},
\end{eqnarray*}
where
\begin{eqnarray*}
\beta_{R} & = & 2.5\,\mathrm{GeV}^{-1},\quad\beta_{P}=3\,\mathrm{GeV}^{-1},\quad\beta_{F}=2.35\,\mathrm{GeV}^{-1}.
\end{eqnarray*}

Now we have enough information to obtain pseudorapidity distributions
of the cross sections of $pp\to c+X$ reaction at different energies.
Experimental data at LHC energies are reported normalized to non-single diffraction
cross section $\sigma_{\mathrm{NSD}}$.
The following parametrization for $\sigma_{\mathrm{NSD}}$ is used in our work:

\begin{eqnarray*}
\sigma_{\mathrm{NSD}}(s) &=& \left[1.76+19.8 \left({s \over \mathrm{GeV^2}}\right)^{0.057} \right]\mathrm{mb}.
\end{eqnarray*}

\begin{figure*}
\begin{centering}
\includegraphics{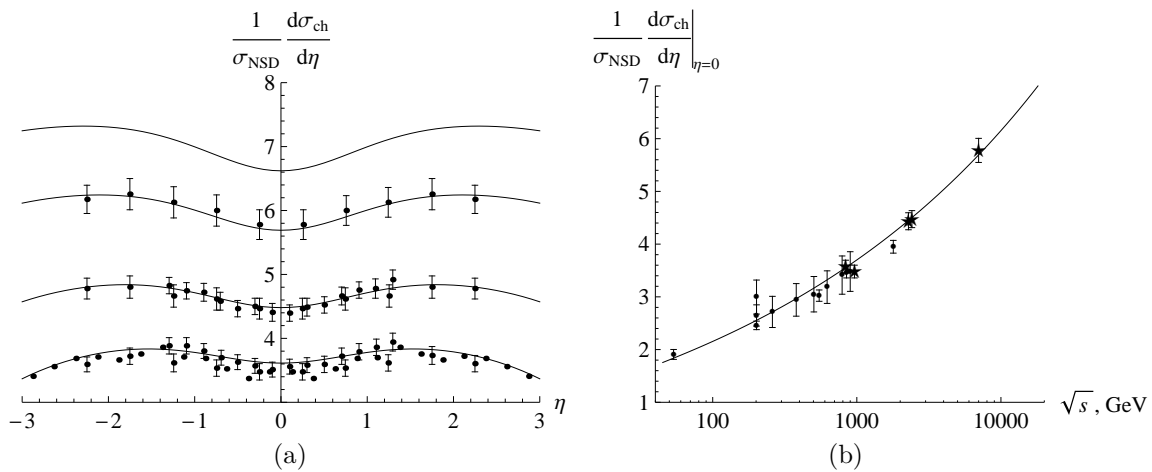}
\par\end{centering}
\caption{(a) --- pseudorapidity distributions of $pp\to c^{\pm}+X$
reaction cross section. Curves in
the figure correspond to energies $\sqrt{s}=0.9$, 2.36, 7 and 14 TeV from
bottom to top; (b) --- energy dependence of differential cross
section of the reaction $pp\to c^{\pm}+X$
at $\eta=0$. Recent values obtained by ALICE and CMS collaborations are marked with stars.
\label{fig:dsdeta}}
\end{figure*}

In Fig. \ref{fig:dsdeta} (a) pseudorapidity
distributions at high energies are shown in comparison with experimental
data taken from the works \cite{Alner:1986xu,Khachatryan:2010us,Aamodt:2010ft}.
In these works distributions for both positively and negatively charged particles
are given, thus they are doubled in comparison with those presented in Fig. \ref{fig:dsdy}.
Energy dependence of differential cross
section of the reaction $pp\to c^{\pm}+X$ at zero rapidity is presented
in Fig. \ref{fig:dsdeta} (b) together with both low-energy data taken from
\cite{Chliap} and recent LHC results \cite{Alner:1986xu,Khachatryan:2010us,Aamodt:2010ft}.
One can clearly see, that
in the framework of Regge phenomenology we get good agreement with the
higher energy experimental data, although these values were not used
in the fitting procedure.

\begin{figure*}
\begin{centering}
\includegraphics{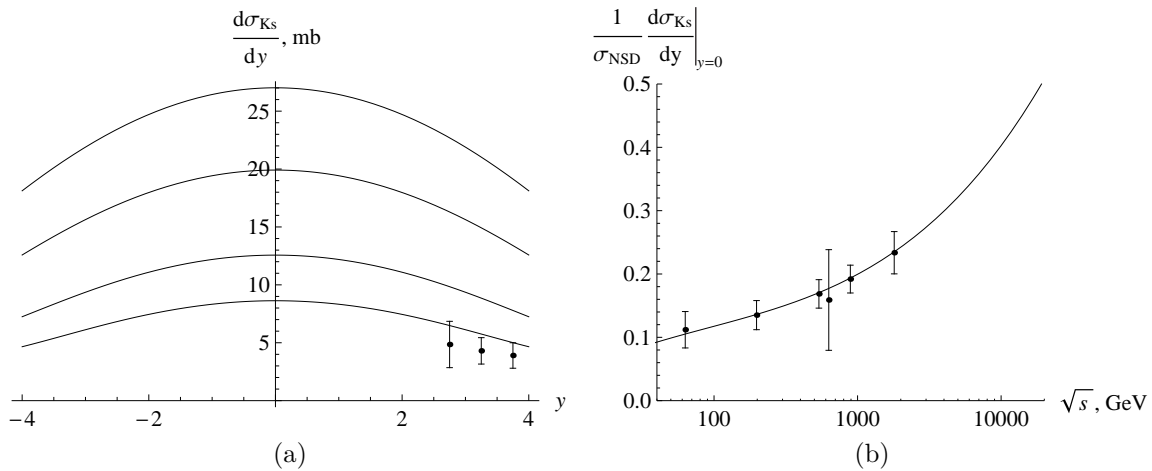}
\par\end{centering}
\caption{(a) --- rapidity distributions of
$pp\to K_{S}+X$ reaction cross section. Curves in
the figure correspond to energies $\sqrt{s}=0.9$, 2.36, 7 and 14 TeV from
bottom to top; (b) --- energy dependence of differential cross
section of the reaction $pp\to K_{S}+X$
at $\eta=0$.
\label{fig:dsHE}}
\end{figure*}

As $K_S$-mesons have definite mass, distributions over rapidity $y$ are experimentally measured.
Recent data from the LHC contains only LHCb results in $2.5<y<4$ region at $\sqrt{s}=0.9$ TeV.
These points together with the Regge model predictions at $\sqrt{s}=0.9$, 2.36, 7 and 14 TeV are
shown in Fig. \ref{fig:dsHE} (a). The agreement within the error bars is achieved but experimental
points are systematically below the theoretical prediction. 
These measurements are reported in units
of cross section as opposed to others, reported in units of multiplicity. Thus the uncertainty
from the normalization cross section may arise.
The rapidity density at $y=0$ have been measured up to 1.8 TeV energy 
\cite{Drijard:1981wg,Alner:1985ra,Ansorge:1987cj,Abe:1989hy}. 
All these measurements are in pretty good agreement with our theoretical predictions.

%

%

\section{Discussion}

The energy dependence of total and inclusive cross sections of particle
production in central region is usually discussed in the framework
of Regge phenomenology. In this approach the amplitude of the process
is written as a sum of reggeonic diagrams: Pomeron and secondary trajectories.
According to perturbative QCD the Pomeron singularity
can be presented as a number of poles in
angular momentum plane with $1\le j\le1+\Delta$.

It was sufficient to consider two
Pomeron poles with the intercepts $\alpha(0)=1$ and
$\alpha(0)=1.17$ to describe low-energy data on proton-proton scattering,
published long before LHC start.
This approach gave both energy dependence of total
cross sections and rapidity distributions at different energy values
\cite{Likhoded91}. We would like to note that these distributions
are strongly correlated with each other.
The constants of Pomeron
interaction with protons and final particles can be determined from a simple
fit of experimental data.
In our article we use low-energy data at $\sqrt{s} \leq 900$ GeV energies to determine
Pomeron coupling constants and predict rapidity and pseudorapidity
distributions of light charged and $K_{S}$-mesons at LHC.
At $\sqrt{s}=0.9$, $2.36$ and 7 TeV energies good agreement with the available
experimental data is observed. We also give predictions of pseudorapidity
distributions of charged particles and $K_{S}$-meson production at
$\sqrt{s}=7$ and $14$ TeV.


The article was financially supported by Russian Foundation for Basic
Research (grant \#10-02-00061a).
The work of A.V. Luchinsky and A.A. Novoselov
was also supported by non-commercial
foundation \textquotedblleft{}Dynasty\textquotedblright{} and the
grant of the president of Russian Federation (grant \#MK-406.2010.2).
A.A. Novoselov was also supported by grant \#MK-406.2010.2.


\begin{thebibliography}{12}
\bibitem{Baker76}M. Baker and K.A. Ter-Martirosyan, Phys.Rep. \textbf{28}
(1976); A.Kaidalov, M.G. Poghosyan, arXiv:0910.2050 (2009).

\bibitem{Lipatov}I.I. Balitsky, L.N. Lipatov, Sov.J.Nucl.Phys. \textbf{28},
822 (1978).

\bibitem{Gribov}L.V. Gribov, E.M. Levin and M.G. Ryskin, Phys.Rep
\textbf{100}, 1 (1983)

\bibitem{Capella}A. Capella, J. Tran Thanh Van, J. Kaplan, Nucl.Phys.
\textbf{B97}, 493 (1975)

\bibitem{Kopelevich} B. Kopelevich, N. Nikolaev and I. Potashnikova,
JINR E2-86-125 (1986)

\bibitem{Petrov02}V.A. Petrov, A.V. Prokudin, Eur.Phys.J \textbf{C23},
135 (2002); hep-ph/0203162.

\bibitem{Kobr}A.K. Likhoded, A.N. Tolstenkov, P.V. Chliapnikov, Yad.Fiz.
\textbf{26} 153 (1977).

\bibitem{Chliap}P.V. Chliapnikov, A.K. Likhoded, V.A. Uvarov, Phys.Lett.
\textbf{B215} 417 (1988), Sov.J.Nucl.Phys.49:1046-1049,1989;

\bibitem{Likhoded91}A.K. Likhoded, O.P. Yushchenko, Int.J.Mod.Phys.
\textbf{A6} 913 (1991).

\bibitem{Abram} V.A. Abramovsky, V.N. Gribov, O.V. Kancheli, Yad.Fiz.
\textbf{18} 595 (1973).

\bibitem{Mueller}A.H. Mueller, Phys.Rev. \textbf{D2}, 2963 (1970);
O.V. Kancheli, Pisma Zh.Eksp.Teor.Fiz. \textbf{11} 397 (1970).

\bibitem{Deta} Vardan Khachatryan et al. (CMS collaboration), JHEP
\textbf{2010}, 1 (2010), arXiv:1002.0621; G. Aad et al. (Atlas collaboration),
Phys.Lett. \textbf{B688}, 21 (2010), arXiv:1003.3124.

\bibitem{Alner:1986xu}
  G.~J.~Alner {\it et al.}  [UA5 Collaboration],
  Z.\ Phys.\  C {\bf 33}, 1 (1986).

\bibitem{Khachatryan:2010us}
  V.~Khachatryan {\it et al.}  [CMS Collaboration],
  Phys.\ Rev.\ Lett.\  {\bf 105}, 022002 (2010)
  [arXiv:1005.3299 [hep-ex]].

\bibitem{Aamodt:2010ft}
  K.~Aamodt {\it et al.}  [ALICE Collaboration],
  Eur.\ Phys.\ J.\  C {\bf 68}, 89 (2010)
  [arXiv:1004.3034 [hep-ex]].
  
\bibitem{Drijard:1981wg}
  D.~Drijard {\it et al.}  [CERN-Dortmund-Heidelberg-Warsaw Collaboration],
  Z.\ Phys.\  C {\bf 12}, 217 (1982).
  
\bibitem{Alner:1985ra}
  G.~J.~Alner {\it et al.}  [UA5 Collaboration],
  Nucl.\ Phys.\  B {\bf 258}, 505 (1985).

\bibitem{Ansorge:1987cj}
  R.~E.~Ansorge {\it et al.}  [UA5 Collaboration],
  Phys.\ Lett.\  B {\bf 199}, 311 (1987).

\bibitem{Abe:1989hy}
  F.~Abe {\it et al.}  [CDF Collaboration],
  Phys.\ Rev.\  D {\bf 40}, 3791 (1989).


\end{thebibliography}
\end{document}